\documentclass[twocolumn,amsmath,amssymb]{revtex4}

\usepackage{graphicx}
\usepackage{dcolumn}
\usepackage{bm}
\newcommand{\etal}{\emph{et al.}}

\begin{document}      

\title{Dissipationless Anomalous Hall Current in the Ferromagnetic Spinel CuCr$_2$Se$_{4-x}$Br$_x$.\footnote{Science {\bf 303}, 1647 
(2004).}} 
\author{Wei-Li Lee$^1$, Satoshi Watauchi$^{2}$\footnote{\emph{Permanent address of S. W. : Center for Crystal Science and Technology, University of 
Yamanashi, 7 Miyamae, Kofu, Yamanashi 400-8511, Japan}}, V. L. Miller$^{2}$, R. J. Cava$^{2,3}$, and N. P. Ong$^{1,3}$\footnote{To whom 
correspondence should be addressed E-mail: npo@princeton.edu}
}      
\affiliation{$^1$Department of Physics, 
$^2$Department of Chemistry, $^3$Princeton Materials Institute, Princeton University, New Jersey 08544, U.S.A.
}
\date{\today}  
\begin{abstract}
In a ferromagnet, an applied electric field $\bf E$ invariably produces an anomalous Hall current ${\bf J}_H$ that flows perpendicular to the plane defined 
by $\bf E$ and $\bf M$ (the magnetization).  For decades, the question whether ${\bf J}_H$ is dissipationless (independent of the scattering rate), has 
been keenly debated without experimental resolution.  In the ferromagnetic spinel CuCr$_2$Se$_{4-x}$Br$_x$, the resistivity $\rho$ (at low temperature) 
may be increased 1000 fold by varying $x$(Br), without degrading the $\bf M$.  We show that ${\bf J}_H/E$ (normalized per carrier, at 5 K) remains 
unchanged throughout.  In addition to resolving the controversy experimentally, our finding has strong bearing on the generation and study of spin-Hall 
currents in bulk samples.

\end{abstract}

\maketitle                   

A major unsettled question in the study of electron transport in a ferromagnet is whether the anomalous Hall current is dissipationless.  In non-magnetic 
metals, the familiar Hall current arises when electrons moving in crossed electric ($\bf E$) and magnetic ($\bf H$) fields are deflected by the Lorentz force.  
However, in a ferromagnet subject to $\bf E$ alone, a large, spontaneous (anomalous) Hall current ${\bf J}_H$ appears transverse to $\bf E$ (in practice, 
a weak $\bf H$ serves to align the magnetic domains) {\it (1,2)}.  Questions regarding the origin of ${\bf J}_H$, and whether it is dissipationless, have 
been keenly debated for decades.  They have emerged anew because of fresh theoretical insights and strong interest in spin currents for spin-based 
electronics.   Here we report measurements in the ferromagnet CuCr$_2$Se$_{4-x}$Br$_x$ which establish that, despite a 100-fold increase in the 
scattering rate from impurities, ${\bf J}_H$ (per carrier) remains constant, implying that it is indeed dissipationless.

In 1954, Karplus and Luttinger (KL){\it (3,4)} proposed a purely quantum-mechanical origin for ${\bf J}_H$.  An electron in the conduction band of a crystal 
lattice spends part of its time in nearby bands because of admixing caused by the (intracell) position operator ${\bf X}$.  In the process, it acquires a 
spin-dependent `anomalous velocity' {\it (5)}.  KL predicted that the Hall current is dissipationless: ${\bf J}_H$ remains constant even as the longitudinal 
current (${\bf J || E}$) is degraded by scattering from added impurities.  A conventional mechanism was later proposed {\it (6)} whereby the anomalous 
Hall effect (AHE) is caused instead by asymmetric scattering of electrons by impurities (skew scattering).  Several authors {\it (7,8,9)} investigated the 
theoretical ramifications of these competing models.  The role of impurities in the anomalous-velocity theory was clarified by Berger's side-jump model {\it 
(7)}.  A careful accounting of various contributions (including side-jump) to the AHE in a semiconductor has been given by Nozi\`eres and Lewiner (NL) 
who derive ${\bf X} = \lambda {\bf k\times S}$, with $\lambda$ the enhanced spin-orbit parameter, ${\bf k}$ the carrier wavevector and $\bf S$ its spin {\it 
(9)}.  In the dc limit, NL obtain the AHE current
\begin{equation}
{\bf J}_H = 2ne^2\lambda {\bf E\times S},
\label{NL}
\end{equation}
where $n$ is the carrier density and $e$ the charge.  As noted, ${\bf J}_H$ is linear in $\bf S$ but independent of the electron 
lifetime $\tau$.

In modern terms, the anomalous velocity term of KL is related to the Berry phase {\it (10)}, and has been applied {\it (11)} to explain the AHE in Mn-doped 
GaAs {\it (12)}.  The close connection of the AHE to the Berry phase has also been explored in novel ferromagnets in which frustration leads to spin 
chirality {\it (13,14,15)}.  In the field of spintronics, several schemes have been proposed to produce a fully polarized spin current in thin-film structures {\it 
(16)}, and in bulk $p$-doped GaAs {\it (17)}.  The AHE is intimately related to these schemes, and our experimental results have bearing on the 
spin-current problem.  

In an AHE experiment {\it (1)}, the observed Hall resistivity is comprised of two terms, 
\begin{equation} 
\rho_{xy} = R_0B + \rho'_{xy},
\label{rhoxy}
\end{equation}
with $B$ the induction field, $R_0$ the ordinary Hall coefficient, and $\rho'_{xy}$ the anomalous Hall resistivity.  A direct test of the dissipationless nature 
of ${\bf J}_H$ is to check whether the anomalous Hall conductivity $\sigma'_H$ (defined as $\rho'_{xy}/\rho^2$) changes as impurities are added to 
increase $1/\tau$ (and $\rho$) {\it (3,7)}.  A dissipationless AHE current implies that $\rho'_{xy} \sim \rho^\alpha$, with $\alpha$ = 2.  By contrast, in the 
skew scattering model, $\alpha = 1$.

Tests based on measurements at high temperatures (77-300 K) yield exponents in the range $\alpha_{exp}$ = 1.4-2.0 {\it (18,19)}.  However, it has been 
argued {\it (20)} that, at high $T$, both models in fact predict $\alpha = 2$, a view supported by detailed calculations {\it (21)}.  To be meaningful, the test 
must be performed in the impurity-scattering regime over a wide range of $\rho$.  Unfortunately, in most ferromagnets, $\rho'_{xy}$ becomes too small to 
measure accurately at low $T$.  Results on $\alpha$ in the impurity-scattering regime are very limited.

The copper-chromium selenide spinel $\rm CuCr_2Se_4$, a metallic ferromagnet with a Curie temperature $T_C\sim$ 430 K, is particularly well-suited 
for testing the AHE.  Substituting Se with Br in $\rm CuCr_2Se_{4-x}Br_x$ decreases the hole density $n_h$ {\it (22)}.  However, because the coupling 
between local moments on Cr is primarily from 90$^\mathrm{o}$ superexchange along the Cr-Se-Cr bonds {\it (23)}, this does not destroy the 
magnetization.  We have grown crystals of $\rm CuCr_2Se_{4-x}Br_x$ by chemical vapor transport [details given in Supporting Online Materials (SOM) {\it 
(24)}].  Increasing $x$ from 0 to 1 in our crystals decreases $n_h$ by a factor of $\sim$30 (Fig. \ref{rho}A), while $T_C$ decreases from 430 K to 230 K.  
The saturated magnetization $M_s$ at 5 K corresponds to a Cr moment that actually increases from $\sim$2.6 to 3 $\mu_B$ (Bohr magneton) (Fig. 
\ref{rho}B).

\begin{figure}[h]			
\includegraphics[width=9cm]{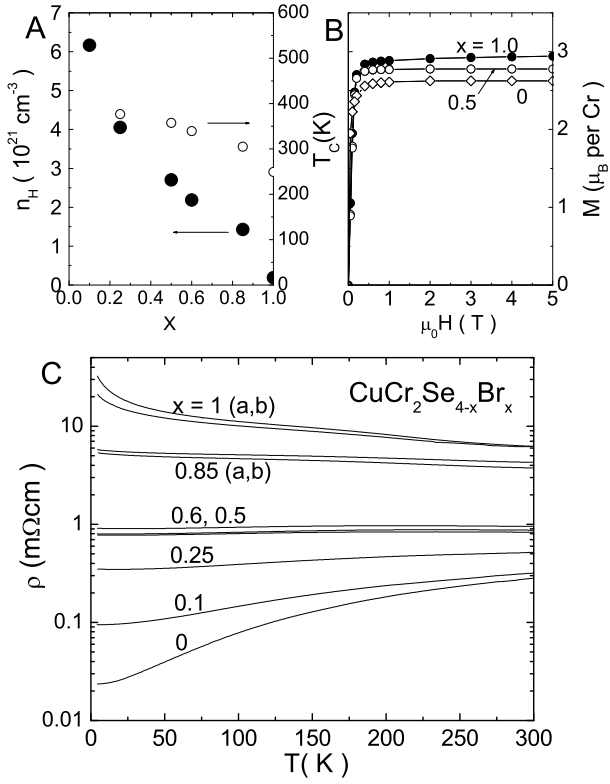}
\caption{\label{rho}  (A) The hole density $n_h$ (solid circles) in $\rm CuCr_2Se_{4-x}Br_x$ vs. $x$ determined from $R_0$ at 400 K (one hole per 
formula unit corresponds to $n_h = 7.2\times 10^{21} \mathrm{cm}^{-3}$).  The Curie temperature $T_C$ is shown as open circles.  (B) Curves of the 
magnetization $M$ vs. $H$ at 5 K in 3 samples ($x$ values indicated).  The saturation value $M_s$ = 3.52, 3.72, 3.95 ($10^5$A/m) for $x$ = 0, 0.5, 1.0, 
respectively.  (C) The resistivity $\rho$ vs. $T$ in 10 samples with Br content $x$ indicated ($a$, $b$ indicate different samples with the same $x$).  
Values of $n_h$ in all samples fall in the metallic regime (for $x = 1$, $n_h = 1.9\times 10^{20}\ \mathrm{cm^{-3}}$).
}
\end{figure}
As shown in Fig. \ref{rho}C, all samples except the ones with $x = 1.0$ lie outside the localization regime.  In the `metallic' regime, the low-$T$ resistivity 
increases by a factor of $\sim$270, as $x$ increases from 0 to 0.85, and is predominantly due to a 70-fold decrease in $\tau$.  The hole density $n_h$ 
decreases by only a factor of 4.  In the localization regime ($x=1.0$), strong disorder causes $\rho$ to rise gradually with decreasing $T$.  We emphasize, 
however, that these samples are not semiconductors ($\rho$ is not thermally activated, and $n_h = 1.9\times 
10^{20}\;\mathrm{cm}^{-3}$ is degenerate). 

The field dependence of the total Hall resistivity (Eq. \ref{rhoxy}) is shown for $x$ = 0.25 (Fig. \ref{rhoxyH}A) and 1.0 (B).  See SOM {\it (24)} for 
measurement details.  The steep increase in $|\rho_{xy}|$ in weak $H$ reflects the rotation of domains into alignment with $\bf H$.  Above the saturation 
field $H_{s}$, when $\rho'_{xy}$ is constant, the small ordinary Hall term $R_0B$ is visible as a linear background {\it (24)}.  As in standard practice, we 
used $R_0$ measured above $T_C$ to find the $n_h$ plotted in Fig. \ref{rho}A.  

\begin{figure}[h]			
\includegraphics[width=9cm]{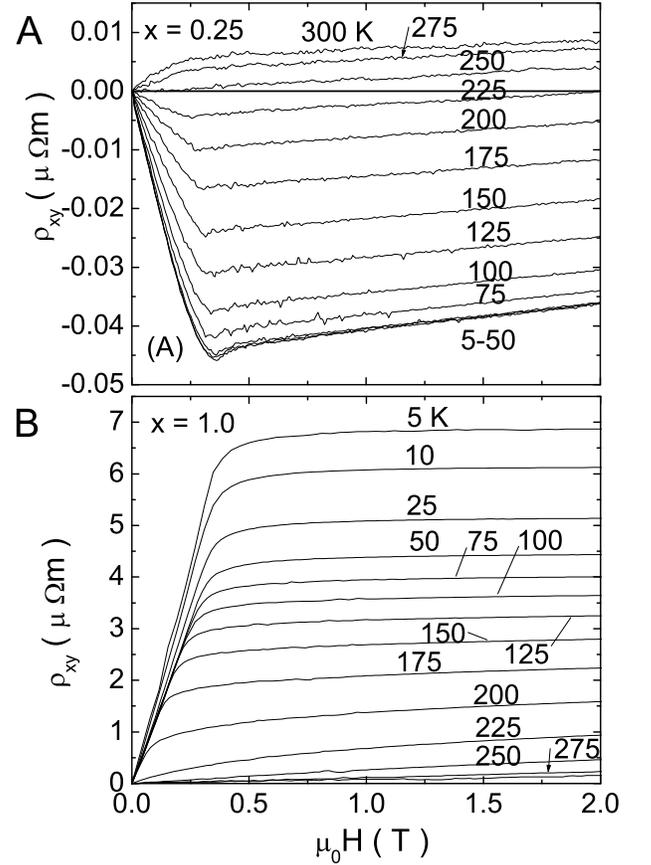}
\caption{\label{rhoxyH}  Curves of the observed Hall resistivity $\rho_{xy} = R_0B + R_s\mu_0M$ vs. $H$ (at temperatures indicated) in $\rm 
CuCr_2Se_{4-x}Br_x$ with $x$ = 0.25 (Panel A) and $x$ = 1.0 (B).  In (A), the anomalous Hall coefficient $R_s$ changes sign below 250 K, becomes 
negative, and saturates to a constant value below 50 K.  However, in (B), $R_s$ is always positive and rises to large values at low $T$ (note difference in 
scale).  
}
\end{figure}
%
%
%
\begin{figure}[h]			
\includegraphics[width=9cm]{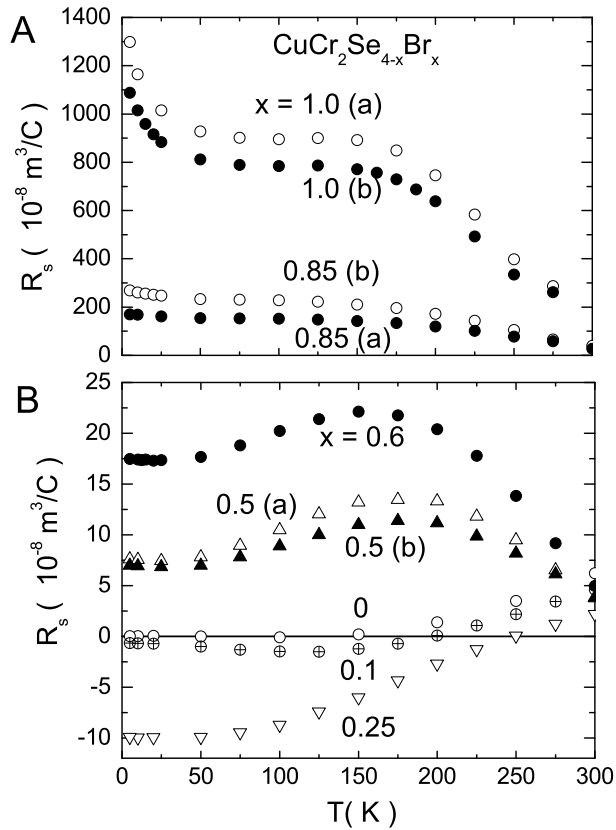}
\caption{\label{RsT}  (A) Values of $R_s$ extracted from the curves of $\rho_{xy}$ and $M$ vs $H$ measured at each $T$ in $\rm CuCr_2Se_{4-x}Br_x$ 
with values of $x$ indicated ($a$ and $b$ refer to different crystals with the same $x$).  The corresponding curves for $x$ = 0, 0.1, 0.25 and 0.5 (2 crystals 
$a$ and $b$) are displayed in Panel (B).  The values of $R_s$ at 5 K are negative at small $x$ ($<0.4$), but as $x$ increases, $R_s$ rapidly rises to 
large positive values. 
}
\end{figure}
By convention, the $T$ dependence of the AHE signal is represented by the anomalous Hall coefficient $R_s(T)$ defined by $\rho'_{xy} = R_s\mu_0M$ 
($\mu_0$ is the vacuum permeability).   By scaling the $\rho'_{xy}$-$H$ curve against the $M$-$H$ curve measured at each $T$, we have determined 
(24) $R_s$ vs. $T$ in each of the samples studied (Fig. \ref{RsT}). The introduction of Br causes the $R_s$ vs. $T$ profiles to change dramatically.  In the 
undoped sample ($x$ = 0), $R_s$ is positive and monotonically decreasing below 360 K, as typical in high-purity ferromagnets (Fig. \ref{RsT}B).  Weak 
doping ($x$ = 0.1) produces a negative shift in $R_s$ and a finite negative value at low $T$.  Increasing the doping to $x$ = 0.25 leads to an $R_s$ 
profile that is large, negative and nearly $T$ independent below 50 K (Fig. \ref{rhoxyH}A).  At mid-range doping and higher ($x\ge$ 0.5), the magnitude of 
$R_s$ increases steeply, but now in the positive direction.  At maximum doping ($x = 1$), the value of $R_s$ at 5 K is very large, corresponding to 
$\rho'_{xy}\sim 700 \ \mu\Omega$cm (Fig. \ref{rhoxyH}A).  

Our focus is on the low-$T$ values of $\rho'_{xy}$ where impurity scattering dominates.  At 5 K, $\rho'_{xy}$ is too small to be resolved in the sample with 
$x$ = 0.  As $x$ increases to 1, the absolute magnitude $|\rho'_{xy}|$ at 5 K increases by over 3 orders of magnitude (from hereon $\rho'_{xy}$ refers to 
the saturated value measured at 2 Teslas or higher).  Significantly, $\rho'_{xy}$ is negative at low doping ($0< x< 0.4$), but becomes positive for $x> 0.5$.  
Initially, the sign-change seemed to suggest to us that there might exist 2 distinct mechanisms for the AHE in this system.  As more samples were studied, 
however, it became apparent that, regardless of the sign, the magnitude $|\rho'_{xy}|$ versus $\rho$ falls on the same curve over several decades (Fig. 
\ref{loglog}), providing strong evidence that the same AHE mechanism occurs in both sign regimes.  We focus first on the magnitude $|\rho'_{xy}|$ vs. 
$\rho$, and discuss the change in sign later.  

It is worth emphasizing that $\sigma'_H$ is proportional to the carrier density $n_h$ (see Eq. \ref{NL}).  For our goal of determining whether the AHE 
current is dissipationless, it is clearly necessary to factor out $n_h$ before comparing $\rho'_{xy}$ against $\rho$.  Hence we divide $|\rho'_{xy}|$ by 
$n_h$.  We refer to $\sigma'_H/n_h$ as the normalized AHE conductivity {\it (24)}.
%
\begin{figure}[h]			
\includegraphics[width=10cm]{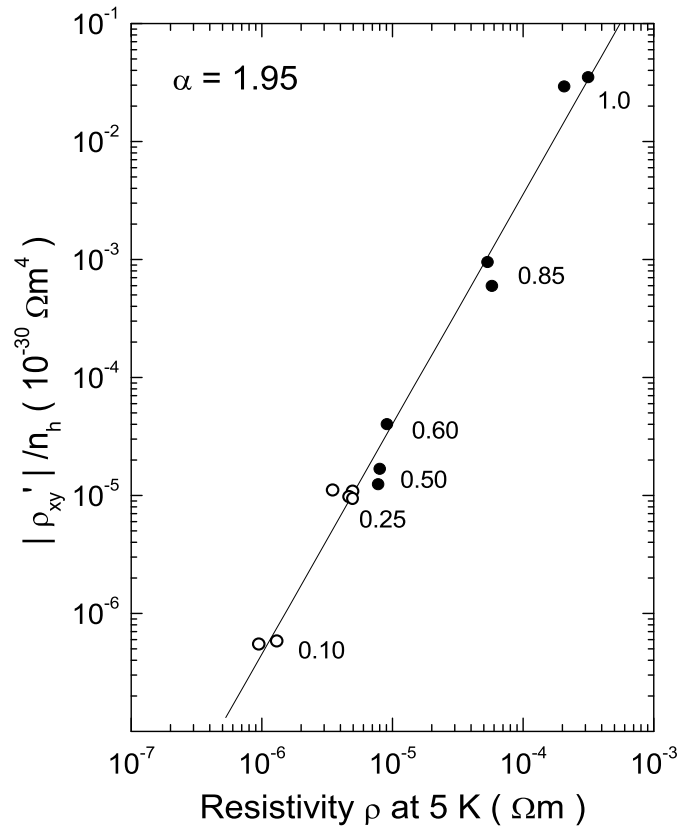}
\caption{\label{loglog}  The normalized quantity $|\rho'_{xy}|/n_h$ versus $\rho$ (at 5 K) in a log-log plot  ($\rho'_{xy}$ is measured at 2 T and 5 K).  The 12 
samples (with doping $x$ indicated) include ones with negative $\rho'_{xy}$ (open circles) and positive $\rho'_{xy}$ (solid).  The undoped sample ($x = 
0$) is not shown because $\rho'_{xy}$ at 5 K is unresolved in our experiment (24).  The least-squares fit gives $\rho'_{xy}/n_h = {\cal A} \rho^\alpha$ with 
${\cal A} = 2.24 \times 10^{-25}$ (SI units) and $\alpha \ = 1.95\pm 0.08$.  
}
\end{figure}

Figure \ref{loglog} displays $|\rho'_{xy}|/n_h$ versus $\rho$ in log-log scale for all samples investigated (except $x = 0$).  Over several decades, the data 
fit well to $|\rho'_{xy}|/n_h = {\cal A}\rho^\alpha$ with $\alpha = 1.95\pm 0.08$ (as $M_s$ is nearly insensitive to $x$, Fig. \ref{loglog} also gives 
$R_s/n_h\sim\rho^2$).  This immediately implies that the normalized AHE conductivity $\sigma_H/n_h$ at 5 K is dissipationless.  Increasing $\rho$ by a 
factor of $\sim 100$ leaves the AHE current per carrier unchanged to our measurement accuracy [see SOM {\it (24)} for a discussion of our resolution].   
As noted, the 2 samples with $x = 1$ are in the localization regime.  The fact that their points also fall on the line implies that the dissipationless nature of 
the normalized AHE current extends beyond the Bloch-state regime (where most AHE theories apply) into the weak localization regime, where much less 
has been done.  This supports recent  theories {\it (10,11,17)} that the anomalous velocity origin is topological in nature, and equally valid in the Bloch and 
localization regimes.

The sign change at $x\sim$ 0.4 is reminiscent of sign changes observed in ferromagnetic alloys (versus composition).  The common feature is that doping 
drives the Fermi energy $\epsilon_F$ across the overlap between two narrow bands derived from distinct transition-metal elements.  In the alloy 
Ni$_{1-x}$Fe$_x$, the band derived from Fe 3$d$ states lies just above the 3$d$ band of Ni.  As $\epsilon_F$ crosses the overlap, $\rho'_{xy}$ changes 
from negative to positive.  Similar sign changes are observed in Au-Fe and Au-Ni alloys.  It has been pointed out {\it (2)} that the spin-orbit parameter 
$\lambda$ in Eq. \ref{NL} changes sign whenever $\epsilon_F$ moves between overlapping narrow bands.  A similar effect is implied in NL's calculation 
{\it (9)}.  Band-structure calculations {\it (25)} on $\rm CuCr_2Se_4$ reveal that $\epsilon_F$ lies in a hole-like band of mostly Cu $3d$ character strongly 
admixed with Cr $3d$ states lying just above.  We infer that, as $\epsilon_F$ rises with increasing Br content, the conduction states acquire more Cr $3d$ 
character at the expense of Cu $3d$, triggering a sign-change in $\lambda$.   The sign change (negative to positive with increasing $x$) is consistent with 
that observed in Ni$_{1-x}$Fe$_x$.  A change-in-sign of the AHE conductivity at band crossings is also described in recent theories 
{\it (10)}.

We now discuss the relevance of our findings to spin-current production.  To produce fully polarized spin currents, it is ideal to use `half metals' 
(ferromagnets in which all conduction electrons are, say, spin-up).  However, only a few examples are known {\it (26)}.  Alternate schemes based on 
elemental ferromagnets have been proposed {\it (16)}.  As evident from Eq. \ref{NL}, anomalous-velocity theories predict that ${\bf J}_H$ depends on the 
carrier spin $\bf S$.  If a beam of electrons with spin populations $n_{\uparrow}$ and $n_{\downarrow}$ enters a region of fixed $\bf M$, the spin-up and 
spin-down electrons are deflected in opposite directions transverse to $\bf E$, just as in the classic Stern-Gerlach experiment.  This results in a Hall 
charge current proportional to the difference between the spin populations, viz. ${\bf J}_H\sim (n_{\uparrow}-n_{\downarrow})$.  More importantly, this also 
produces a fully polarized spin Hall current ${\bf J}_s$ proportional to the sum $(n_{\uparrow}+n_{\downarrow})$.  Hence, in a ferromagnet that is not a half 
metal, the spin Hall current is fully polarized according to these theories.  By contrast, in skew-scattering theories, ${\bf J}_H$ depends on the direction of 
the local moment ${\bf m}_{i}$ on the impurity but not the spin of the incident electron (i.e. both $J_H$ and 
$J_s\sim(n_{\uparrow}-n_{\downarrow})$).  

In confirming that the normalized AHE current is dissipationless over a multi-decade change in $\rho$, we verify a specific prediction of the 
anomalous-velocity theories and resolve a key controversy in ferromagnets.  The implication is then that fully polarized spin-Hall currents are readily 
generated in ferromagnets (at low $T$) by simply applying $\bf E$.  While this realization does not solve the conductivity-mismatch problem at interfaces 
{\it (27)}, it may greatly expand the scope of experiments on the properties of spin currents.   

\newpage
\title{Supplementary Online Material for Dissipationless Anomalous Hall Current in the Ferromagnetic Spinel 
CuCr$_2$Se$_{4-x}$Br$_x$.} 
\author{Wei-Li Lee$^1$, Satoshi Watauchi$^{2}$\footnote{\emph{Permanent address of S. W. : Center for Crystal Science and Technology, University of 
Yamanashi, 7 Miyamae, Kofu, Yamanashi 400-8511, Japan}}, V. L. Miller$^{2}$, R. J. Cava$^{2,3}$, and N. P. Ong$^{1,3}$\footnote{To whom 
correspondence should be addressed E-mail: npo@princeton.edu}
}      
\affiliation{$^1$Department of Physics, 
$^2$Department of Chemistry, $^3$Princeton Materials Institute, Princeton University, New Jersey 08544, U.S.A.
}

\maketitle                   

{\bf Supporting Online Material}\\
{\bf Materials and Methods}\newline
Mixtures of Cu, Cr, Se and CuBr$_2$ powder were heated at 550-600$\rm ^o$C for 20 h in evacuated sealed quartz tubes.  The reactants ($\sim$3 g) 
were pulverized and sealed in quartz tubes (1.4 $\times$ 15 cm) with iodine ($\sim$0.3 g) as the transport gas for the crystal growth.  The temperature 
gradient was fixed at 6.7$\rm ^o$C/cm (ends at $\sim$870 and $\sim$770 $\rm ^o$C ) during the 2-week period of growth.  The Br content was analyzed 
by EDX spectroscopy.  For the Hall measurements, samples of typical size 1.5 $\times$0.5$\times$0.1 mm$^3$ were cut from as-grown 
crystals.  

\noindent
{\bf Measurements and analysis}\newline
On each crystal, microspots of indium solder were used to attach a pair of current $I$ leads and 2 pairs of transverse voltage $V$ leads in the standard 
Hall-bar geometry.  The Hall voltage $V_{xy}$ was measured (by ac phase-sensitive detection) in a field swept from -2 T to +2 T at the rate 5 mT/s.  The 
value of $I$ equals 1 mA for samples with $0\le x < 0.5$, and 0.1 mA for $0.5\le x \le 1$.  The Hall resistivity is calculated from the antisymmetric part of the 
Hall voltage, viz. $\rho_{xy} = [V_{xy}(H)-V_{xy}(-H)]/2I$ to remove the ``IR'' drop from Hall lead misalignment (typically, the ``IR'' background, which is 
symmetric in $H$, accounts for less than 5$\%$ of the recorded Hall voltage).  In most samples, we checked for current uniformity by comparing the Hall 
signals from the 2 pairs of Hall voltage leads.  The two signals agree to better than 5$\%$.  The largest source of error arises from measurements of the 
crystal thickness and the finite size of the microdot contacts (which affect the effective length and width).  We estimate that these geometric errors 
contribute $\pm 7\;\%$ to the uncertainty in the absolute values of $\rho_{xy}$ and $\rho$. 

\begin{figure}[h]			
\includegraphics[width=9cm]{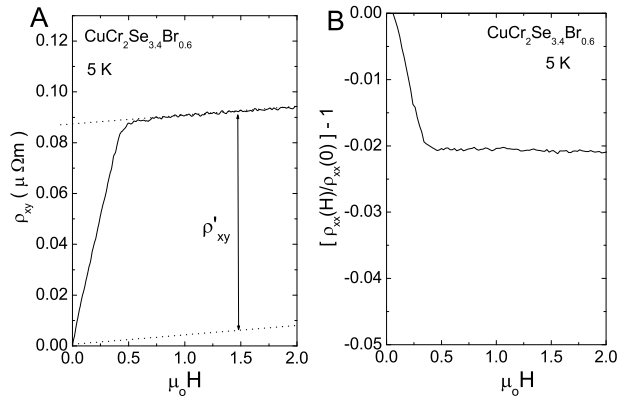}
\caption{\label{S1}
(Panel A) Trace of $\rho_{xy}$ vs. $H$ in CuCr$_2$Se$_{4-x}$Br$_x$ ($x = 0.6$) obtained by antisymmetrizing the recorded Hall voltage.  The linear 
background (broken line) is the term $R_0B$ in Eq. \ref{R0}.  The AHE resistivity $\rho'_{xy}$ is indicated by double arrow.  Panel B shows the weak 
magnetoresistance of the same sample at 5 K.  The fractional decrease $\rho(H)/\rho(0)-1$ is $\sim 2\%$.} 
\end{figure}

Figure \ref{S1} shows a `raw'  trace of the observed Hall resistivity vs. $H$ at $T$ = 5 K in a crystal with $x$ = 0.6.  The trace is consistent with the 
equation
\begin{equation}
\rho_{xy} = R_0 B + \rho'_{xy}, \quad\quad (\rho'_{xy} = R_s \mu_0 M) 
\label{R0}
\end{equation}     
where $R_0$ is the ordinary Hall coefficient and $R_s$ the AHE coefficient.  The induction field ${\bf B} = \mu_0[{\bf H} + (1-N){\bf M}]$ is the sum of the 
applied field $\bf H$ and the magnetization $\bf M$ reduced by the factor $(1-N)$, where $N$ is the demagnetization factor ($N\sim 0.7$-$0.95$ with $\bf 
H$ normal to the plate-like crystals). 

The anomalous Hall resistivity $\rho'_{xy}$ is the main quantity of interest in Eq. \ref{R0}.  As shown in Fig. \ref{S1}, it is obtained by subtracting the linear 
background (representing $R_0B$) from the measured $\rho_{xy}$ (note that the demagnetization factor $N$ is irrelevant to $\rho'_{xy}$).  After 
$\rho'_{xy}$ is obtained, it is found to match (up to a scale factor) the profile of $M(T)$ vs. $H$ measured at the same $T$.  The scale factor gives the 
AHE coefficient $R_s(T)$ which is plotted against $T$ in Fig. 3 (main text) for each sample.  

Below $T_C$, the ordinary Hall coefficient $R_0$ is technically difficult to determine from the linear background term $R_0B$ because it is enhanced by 
the term $M(1-N)$.  In addition, a significant $H$-linear term may arise from the so-called ``paraprocess" susceptibility $\partial M/\partial H$ (which is 
amplified if $R_s\gg R_0$).  These contributions are hard to estimate accurately (\emph{Ref. 1, p. 158}).  As in standard practice, we have measured 
$R_0$ at temperatures near $T_C$ or above it and assumed that the hole density $n_h= (eR_0)^{-1}$ is $T$ independent.   We note that the hole density 
determined is in nominal agreement with the chemical argument that each Br removes one hole per formula unit.

\noindent\emph{AHE conductivity}\newline
The total Hall conductivity $\sigma^{tot}_{xy}$ is the sum of the ordinary Hall conductivity ($\sigma^0_{xy}$) and the AHE conductivity ($\sigma'_{xy}$).  By 
matrix inversion, we then have 
\begin{equation}
\sigma^{tot}_{xy} = \sigma^0_{xy} + \sigma'_{xy} = \frac{\rho_{xy}}{[\rho^2 + \rho_{xy}^2]} \simeq \frac{\rho_{xy}}{\rho^2}.
\label{sigma}
\end{equation}
In our samples, the Hall angle ratio $\tan\theta_H = \rho_{xy}/\rho$ varies (at 5 K) from $5\times 10^{-3}$ at $x = 0.1$ to $2\times 10^{-2}$ at $x = 1.0$.  
Hence at any $x$, the correction $(\rho_{xy}/\rho)^2$ is negligibly small.  This justifies its neglection in the second step (this seems the case for all 
published Hall results on ferromagnets).   Identifying $\sigma^0_{xy}$ with $R_0B/\rho^2$, and comparing Eqs. \ref{R0} with \ref{sigma}, we obtain for the 
AHE conductivity 
\begin{equation}
\sigma'_{xy} = \rho'_{xy}/\rho^2.
\label{sxy}
\end{equation}
By Eq. \ref{sxy}, if $|\rho'_{xy}|/n_h$ is proportional to $\rho^2$, the magnitude of the normalized AHE conductivity $|\sigma'_{xy}|/n_h$ is independent of 
$\tau$.

\noindent\emph{Magnetoresistance}\newline  
At the saturation field when domains become aligned, the elimination of domain walls leads to a slight reduction in carrier scattering which is observed as 
a negative magnetoresistance (MR).  This MR does not affect the extraction of $\rho'_{xy}$ (since it is symmetric in $H$), but it comes in when we 
calculate $\sigma'_{xy}$ from Eq. \ref{sxy}.  However, in all our samples, the negative MR is a 1-2 $\%$ effect (Fig. \ref{S1}B).  Inclusion of this small MR 
correction to $\rho$ leads to a change that is unresolvable in Fig. 4.

\noindent\emph{Resolution}\newline
For samples with $x\ge 0.5$, the relatively large AHE resistivity $\rho'_{xy}$ at 5 K may be measured with reasonably high accuracy.  In the low-$x$ limit, 
however, the steep decrease of $\rho$ (at 5 K) causes $|\rho_{xy}|$ to fall very rapidly towards zero.  We discuss our resolution in 
this limit.
\begin{figure}[h]			
\includegraphics[width=9cm]{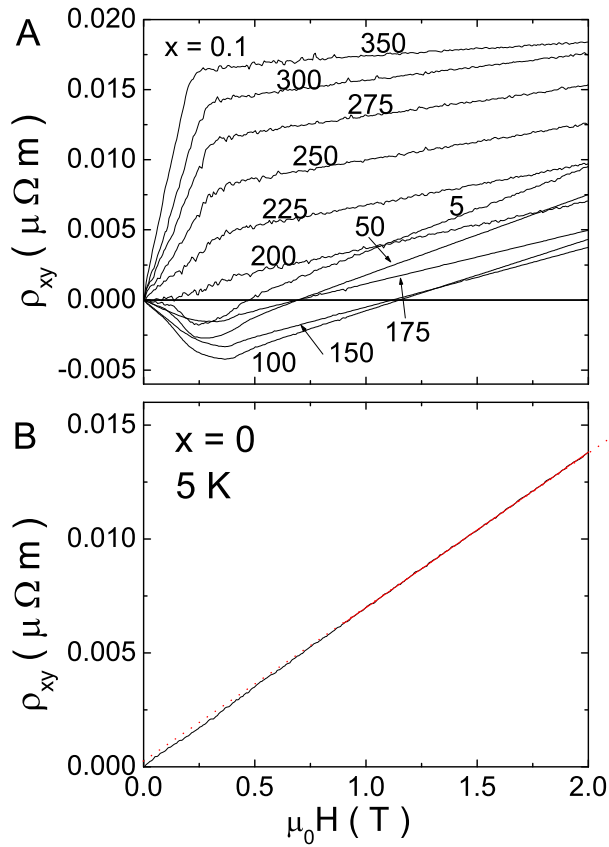}
\caption{\label{S2}
(Panel A)  Curves of the observed Hall resistivity $\rho_{xy} = R_0B + R_s\mu_0M$ vs. $H$ 
in $\rm CuCr_2Se_{4-x}Br_x$ with $x$ = 0.10.  $R_s$ changes from positive to negative as $T$ falls below 180 K, and rapidly decreases in magnitude.  
However, it remains clearly resolvable at 5 K.  Panel B shows the curve of $\rho_{xy}$ vs. $H$ in the sample with $x = 0$ at 5 K.  The close fit to a straight 
line (red dashed line) implies that $\rho'_{xy}$ is below our resolution in this sample.}
\end{figure}
Figure \ref{S2} compares the curves of $\rho_{xy}$ measured in a sample with $x$ = 0.10 (Panel A) with one at $x$ = 0 (B).  In Panel A, the AHE 
component $\rho'_{xy}$, with its characteristic `knee' profile is readily distinguished at high $T$.  As $T$ decreases below 175 K, $R_s$ becomes 
negative.  Further, as $T$ decreases, the slope of the background term $R_0B$ increases noticeably, an effect commonly observed in ferromagnets.  
Despite the large $R_0B$ term at low $T$, the knee profile of $\rho'_{xy}$ remains clearly resolved down to 5 K.  This is the lowest data point displayed in 
Fig. \ref{loglog}  

The undoped sample ($x = 0$) provides a quantitative test of our resolution.  Panel B shows the raw trace of its $\rho_{xy}$ vs. $H$ at 5 K, in which the 
AHE signall is not resolved.  The data above 0.7 T fit very closely to a straight line (dashed red line).  From the small intercept of the fit at $H = 0$, we 
estimate an upper bound ($|\rho'_{xy}|< 2\times 10^{-10}\; \Omega$m) for the AHE signal.  Because the AHE signal at 5 K is unresolved, the $x = 0$ 
sample is not displayed in Fig. 4.  Significantly, however, the upper bound (which implies that $|\rho'_{xy}|/n_h < 2.8\times 10^{-38}$ 
$\Omega\mathrm{m}^4$) is consistent with the best linear fit in Fig. 4.  The value of $|\rho'_{xy}|/n_h$ predicted by extrapolating the straight-line fit to the 
value of $\rho\; ( = 24\;\mu\Omega\mathrm{cm}$ at 5 K) in this sample would fall slightly below our resolution.  This illustrates the problem of measuring 
$\rho'_{xy} = \sigma'_{xy}\rho^2$ in high-purity ferromagnets at low $T$.  Even if $\sigma'_{xy}$ is sizeable, $\rho$ may be too small to render $\rho'_{xy}$ 
observable.

\newpage
\noindent{\bf References and Notes}\newline
1. \emph{The Hall Effect in Metals and Alloys}, ed. Colin Hurd (Plenum, New York 1972) p. 153.\newline
2. L. Berger, G. Bergmann, \emph{The Hall Effect and Its Applications}, ed. C. L. Chien, C. R. Westgate (Plenum, New York 1980), p. 
55.  \newline
3. R. Karplus, J. M. Luttinger, {\it Phys. Rev.} {\bf 95}, 1154 (1954).\newline
4. J. M. Luttinger, {\it Phys. Rev.} {\bf 112}, 739 (1958).\newline
5. E. N. Adams, E. I. Blount, {\it J. Phys. Chem. Solids} {\bf 10}, 286 (1959).\newline
6. J. Smit, {\it Physica (Amsterdam)} {21}, 877 (1955).\newline
7. L. Berger, {\it Phys. Rev. B}  {\bf 2}, 4559 (1970).\newline
8. S. K. Lyo, T. Holstein, {\it Phys. Rev. B} {\bf 9}, 2412 (1974).\newline
9. P. Nozi\`{e}res, C. Lewiner, {\it J. Phys. (France)} {\bf 34}, 901 (1973).\newline
10. M. Onoda, N. Nagaosa, {\it J. Phys. Soc. Jpn.} {\bf 71}, 19 (2002).\newline
11. T. Jungwirth, Qian Niu, A. H. MacDonald, {\it Phys. Rev. Lett.} {\bf 88}, 207208 (2002).\newline
12. H. Ohno, {\it Science} {\bf 281}, 1660 (1998).\newline
13. P. Matl \etal, {\it Phys. Rev. B} {\bf 57}, 10248 (1998).\newline
14. J. Ye \etal, {\it Phys. Rev. Lett.} {\bf 83}, 3737 (1999). \newline
15.  Y. Taguchi \etal, {\it Science}  {\bf 291}, 2573 (2001).\newline
16. J. E. Hirsch, {\it Phys. Rev. Lett.} {\bf 83}, 1834 (1999).\newline
17. S. Murakami, N. Nagaosa, S. C. Zhang, {\it Science} {\bf 301}, 1348 (2003).\newline
18. C. Kooi, {\it Phys. Rev.} {\bf 95}, 843 (1954).\newline
19. W. Jellinghaus, M. P. DeAndres, {\it Ann. Physik} {\bf 7}, 189 (1961).\newline 
20. J. Smit, {\it Phys. Rev. B} {\bf 8}, 2349 (1973).\newline
21. S. K. Lyo, {\it Phys. Rev. B} {\bf 8}, 1185 (1973).\newline
22. K. Miyatani \etal, {\it J. Phys. Chem. Solids} {\bf 32}, 1429 (1971).\newline
23. J. B. Goodenough, {\it J. Phys. Chem. Solids} {\bf 30}, 261 (1969).\newline
24. See Supporting Online Material.\newline
25. F. Ogata, T. Hamajima, T. Kambara, K. I. Goondaira, {\it J. Phys. C.} {\bf 15}, 3483 (1982).\newline
26. R. J. Soulen Jr. \etal, {\it Science} {\bf 282}, 85 (1998).\newline
27. G. Schmidt, D. Ferrand, L. W. Molenkamp, A. T.  Fiolip, B. J. van Wees, {\it Phys. Rev. B} {\bf 62}, R4790 (2000).\newline
28. We acknowledge support from a MRSEC grant (DMR 0213706) from the U. S. National Science Foundation.\newline

\noindent{\bf Supporting Online Material}\newline
www.sciencemag.org\newline
Materials and Methods\newline

\end{document}